\documentclass[aps,pre,reprint,superscriptaddress,longbibliography,nobalancelastpage,groupedaddress,floatfix]{revtex4-2}
\usepackage{graphicx}
\usepackage{dcolumn}
\usepackage{amsmath,amssymb}
\usepackage{latexsym}
\usepackage{mathrsfs}
\usepackage[usenames,dvipsnames]{color}
\usepackage{newtxtext}
\usepackage[slantedGreek]{newtxmath}
\usepackage{hyperref}
\usepackage{xcolor}
\usepackage[compact]{titlesec}
\usepackage{orcidlink}

\usepackage[normalem]{ulem}
\newcommand{\be}{\begin{equation}} 
\newcommand{\ee}{\end{equation}} 
\newcommand{\bea}{\begin{eqnarray}} 
\newcommand{\eea}{\end{eqnarray}} 

\newcommand{\figref}[2]{[Fig.~\hyperref[#1]{\ref*{#1}(#2)}]}
\newcommand{\figrefi}[2]{[Fig.~\hyperref[#1]{\ref*{#1}(#2)}, inset]}
\newcommand{\textfigref}[2]{Fig.~\hyperref[#1]{\ref*{#1}(#2)}}
\newcommand{\textfigureref}[2]{Figure~\hyperref[#1]{\ref*{#1}(#2)}}
\newcommand{\wholefigref}[1]{(Fig.~\ref{#1})}

\newcommand{\textwholefigref}[1]{Fig.~\ref{#1}}
\newcommand{\citeref}[1]{Ref.~\cite{#1}}
\newcommand{\figrefp}[2]{\hyperref[#1]{\ref*{#1}(#2)}}

\definecolor{linkcolor}{HTML}{223096}
\hypersetup{colorlinks,allcolors=linkcolor}

\bibpunct{\textcolor{linkcolor}{[}}{\textcolor{linkcolor}{]}}{\textcolor{linkcolor}{,}}{n}{}{;}
\renewcommand{\eqref}[1]{\hyperref[#1]{(\ref*{#1})}}
\DeclareMathAlphabet{\mathcal}{OMS}{cmsy}{m}{n}

\renewcommand{\epsilon}{\varepsilon}
\newcommand{\citeappendix}[1]{Appendix~\ref{#1}}

\renewcommand{\pi}{\uppi}

\begin{document}
\title{Discontinuous buckling from cytoplasmic mechanics in vertex models of epithelia}

\author{Chandraniva \surname{Guha Ray}\,\orcidlink{0009-0002-8047-7441}}
\affiliation{Max Planck Institute for the Physics of Complex Systems, N\"othnitzer Stra\ss e 38, 01187 Dresden, Germany}
\affiliation{\smash{Max Planck Institute of Molecular Cell Biology and Genetics, Pfotenhauerstra\ss e 108, 01307 Dresden, Germany}}
\affiliation{Center for Systems Biology Dresden, Pfotenhauerstra\ss e 108, 01307 Dresden, Germany}
\author{Pierre A. Haas\,\orcidlink{0000-0002-6663-0393}}
\email[Contact author: ]{haas@pks.mpg.de}
\affiliation{Max Planck Institute for the Physics of Complex Systems, N\"othnitzer Stra\ss e 38, 01187 Dresden, Germany}
\affiliation{\smash{Max Planck Institute of Molecular Cell Biology and Genetics, Pfotenhauerstra\ss e 108, 01307 Dresden, Germany}}
\affiliation{Center for Systems Biology Dresden, Pfotenhauerstra\ss e 108, 01307 Dresden, Germany}
\date{\today}
\begin{abstract}
The cell cortex and cell adhesion contribute to the mechanics of epithelial cells and hence tissues. Vertex models of epithelia represent the cell cortex and cell adhesion, but these models mostly ignore the other subcellular structures that compose the cytoplasm surrounded by the cortex. Here, we reveal the mechanical consequences of these structures in a minimal vertex model of an epithelial monolayer, in which the single-cell energy includes a cytoplasmic contribution that represents, e.g., the cost of confinement of cytosolic polymers. Strikingly, our simulations show that the buckling transition of the monolayer under compression becomes discontinuous in an intermediate range of values of the associated cytoplasmic modulus. By asymptotic solution of this buckling problem in the continuum limit of the model, we explain this reentrant behaviour in terms of the tricritical points of a sixth-order Landau theory. Our results thus underline how cell-scale mechanics beyond the cortex and cell adhesion can change emergent, tissue-scale behaviour even qualitatively. 
\end{abstract}\maketitle

\section{\uppercase{Introduction}}
\begin{figure}[b]
    \centering
    \includegraphics[width=8.6cm]{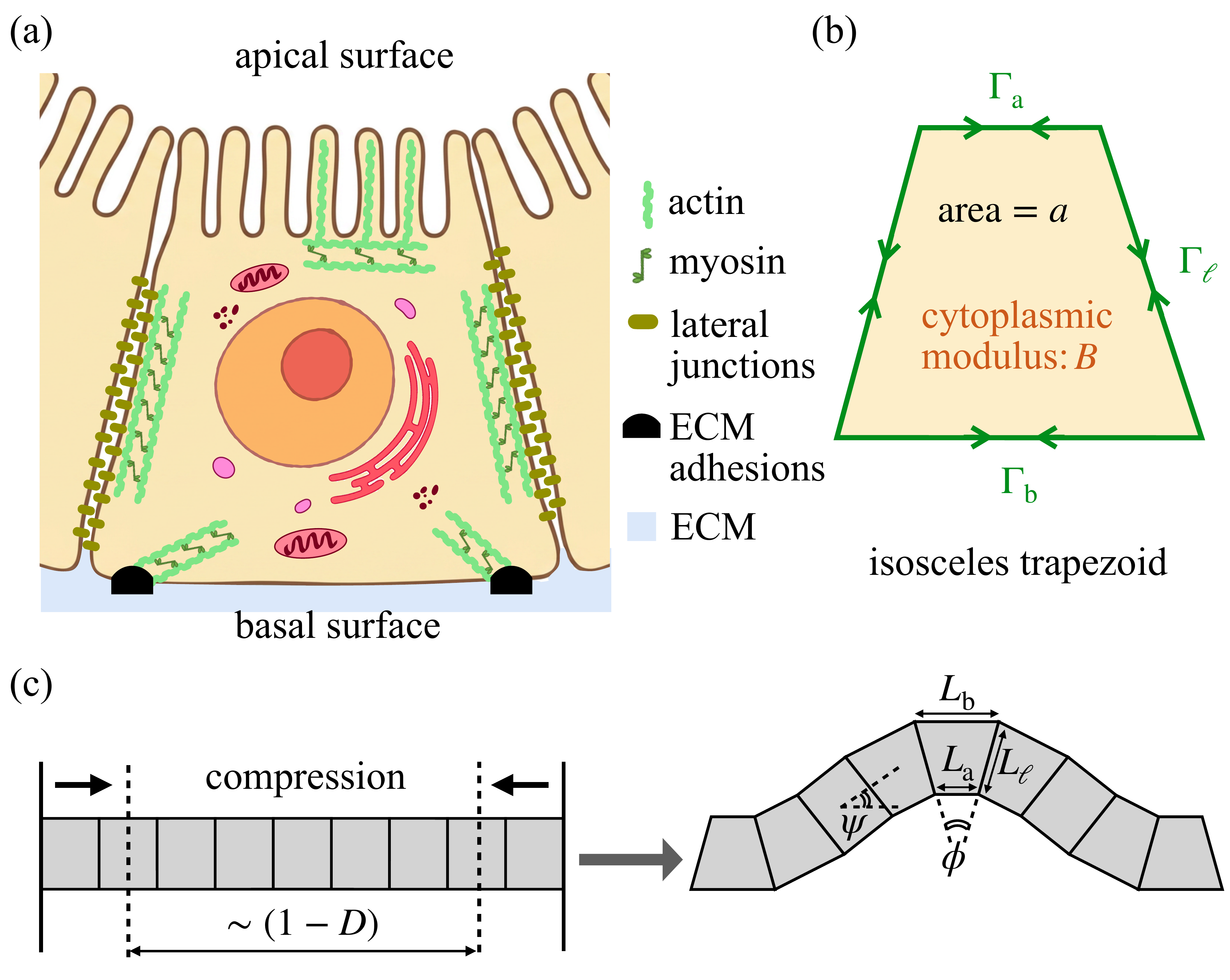}
    \caption{Epithelial cell biology and mechanics. (a)~Schematic representation of an epithelial cell, based on Ref.~\cite{guharay26}. A~contractile actomyosin cortex underlies the apical, basal, and lateral cell surfaces. On the lateral and basal sides, the cells adhere to their neighbours and to the surrounding extracellular matrix. The cytosol enclosed by the cell membranes is a dense polymeric soup suspending the nucleus and other organelles. (b)~Minimal vertex model: each cell is an isosceles trapezoid of fixed area $a$ and respective apical, basal, and lateral surface tensions $\Gamma_\text{a}$,~$\Gamma_\text{b}$, and $\Gamma_\ell$. Cytoplasmic mechanics are associated with a modulus~$B$. (c) Buckling of a tissue under external compression by a relative amount $D$: as the tissue buckles, the initially rectangular cells become trapezoidal. The tangent angle to the buckled midline is $\psi$, and the lateral cell sides are at an angle $\phi$ to each other. The apical, basal, and lateral cell sides are $L_\text{a},L_\text{b},L_\ell$, respectively.}
    \label{fig1}
\end{figure}

The mechanics of cell surfaces~\cite{lecuit07} contribute to the shapes of epithelial cells~\figref{fig1}{a} and hence of the epithelial tissues that are found in all animals. These surface mechanics are defined by the contractile actomyosin cortex that underlies all cell surfaces as well as cell adhesion to neighbouring cells and to the surrounding extracellular matrix (ECM)~\cite{lecuit07,lecuit11}. Many epithelial cells display apicobasal polarisation~\cite{buckley22} with distinct apical, basal, and lateral surface domains \figref{fig1}{a}, so that the basal surfaces face the ECM. The different cell surfaces therefore have, in general, different mechanical properties~\cite{guyomar24}, combining positive surface tensions from the actomyosin cortex and, on the lateral and basal surfaces, negative surface tensions from cell-cell and cell-ECM adhesion~\cite{hannezo14,guharay26}.

Vertex models of epithelia~\cite{hannezo14,derganc09,fletcher14,fletcher16,alt17,krajnc18,rozman20,rozman21} directly encode these cell surface mechanics into the single-cell energy of their polygonal or polyhedral cells. In particular, lateral vertex models or differential-tension models~\cite{hocevar12,krajnc13,storgel16} focus on a two-dimensional apicobasal cross section. In their minimal incarnation~\figref{fig1}{b}, each cell is an isosceles trapezoid of fixed area $a$ and apical, basal, and sides of lengths $L_\text{a}$, $L_\text{b}$, and $L_\ell$, respectively, with corresponding line tensions $\Gamma_\text{a}$, $\Gamma_\text{b}$, and $\Gamma_\ell$. The single-cell energy is thus
\begin{align}
e=\Gamma_\text{a}L_\text{a}+\Gamma_\text{b}L_\text{b}+\Gamma_\ell L_\ell.\label{eq:difften}
\end{align}
The coarse-grained, continuum picture~\cite{krajnc15,haas19,andrensek23,guharay25,andrensek26} that emerges from these differential-tension models at the tissue scale highlights parallels, but also fundamental differences to elastic models: for example, we have recently discovered that the buckling amplitude of a compressed monolayer can decrease with increasing compression in these models~\cite{guharay25}. This unbuckling bifurcation results from the geometric singularity, absent in elastic theories, associated with triangular cell shapes that, biologically, represent cell constriction~\cite{sawyer10,keller11}.

The mechanical properties of epithelial cells and tissues have, of course, beyond these surface terms, additional contributions from the bulk of the cytoplasm~\figref{fig1}{a}. For example, the nucleus, significantly stiffer than the cytoskeleton~\cite{lammerding11}, plays important mechanical roles~\cite{kalukula22}, including in epithelial morphogenesis~\cite{ambrosini19,deleeuw24,roby25}. Recent work~\cite{feng18,phuyal23} has begun to elucidate the function of mechanical forces on organelles, and conversely, it is likely that organelles contribute to defining the mechanical properties of the cytoplasm. Similarly, the cytosol is a dense soup of cytoskeletal and other proteins, such as intermediate filaments~\cite{koster15}, that are expected to resist cell shape deformations.

However, these subcellular structures have largely been ignored in models of epithelial tissues: we are aware of only one study~\cite{hannezo14} that included a cytoplasmic contribution that models these structures explicitly in a vertex model~\footnote{Some vertex models~\cite{honda04,hannezo14,zhang22,villeneuve24,sarkar26} include a volume elasticity term, rather than imposing cell incompressibility. This bulk contribution is perhaps not a model of a specific cytoplasmic structure.}.\nocite{honda04,zhang22,villeneuve24,sarkar26} The emergent role of cytoplasmic mechanics at the scale of epithelial tissues has thus remained unexplored.

Here, we study a minimal model of an epithelial monolayer that extends the minimal differential-model \eqref{eq:difften} by including a cytoplasmic contribution in the spirit of Ref.~\cite{hannezo14}. We combine numerical simulations and analytical calculations of the buckling of the monolayer under external compression~\figref{fig1}{c} to reveal how these bulk mechanics can lead to even qualitative changes of the buckling behaviour of the monolayer.

\section{\uppercase{Vertex model of cytoplasmic mechanics}}
We introduce a minimal vertex model in which, as in Refs.~\cite{haas19,guharay25}, each cell is an isosceles trapezoid of area $a$ \figref{fig1}{b}. The apical, basal, and lateral cell sides have respective lengths $L_\text{a},L_\text{b},L_\ell$~\figref{fig1}{c}. The corresponding apical, basal, and lateral surface tensions are $\Gamma_\text{a},\Gamma_\text{b},\Gamma_\ell$, respectively. As in Ref.~\cite{guharay25}, we assume the apical and basal tensions to be equal, $\Gamma_\text{a}=\Gamma_\text{b}=\Gamma$. This removes the spontaneous wrinkling instability~\cite{krajnc13,andrensek23} that, for sufficiently large $|\Gamma_\text{a}-\Gamma_\text{b}|$, arises in less minimal differential-tension models and that we are not interested in here. 

We now add a cytoplasmic contribution to the surface energy~\eqref{eq:difften}. We choose to focus on a specific contribution, associated with the energetic cost of confinement of cytoplasmic polymers to the cell shape. This has previously been modelled in Ref.~\cite{hannezo14}: confinement to a length $L\in\{L_\text{a},L_\text{b},L_\ell\}$ has an energetic cost proportional to $L^{-2}$ \cite{* [{Reference~\cite{hannezo14} discusses the more general case of an energetic cost proportional to $L^{-\xi}$, too, noting that the exponent $\xi=2$ corresponds to the confinement of Gaussian polymers [see, e.g., }] [{], but that a different exponent, $\xi=7/2$, has been suggested for semiflexible polymers by }] degennes, *claessens06}. The single-cell energy therefore becomes
\begin{align}
e=\Gamma(L_\text{a}+L_\text{b})+\Gamma_\ell L_\ell+B\biggl(\dfrac{1}{L_\text{a}^2}+\dfrac{1}{L_\text{b}^2}+\dfrac{1}{L_\ell^2}\biggr),\label{eq:ecell}
\end{align}
in which $L_\text{a},L_\text{b},L_\ell$ are related by the constraint of area conservation. We will refer to $B$ as the cytoplasmic modulus.

\subsection{Nondimensionalisation}
As in Refs.~\cite{haas19,guharay25}, we express the single-cell geometry in terms of the angle $\phi$ that the lateral cell sides make with each other \figref{fig1}{c}. Thus
\begin{align}
&L_\text{a}=\dfrac{a}{L_\ell}\sec{\phi}+L_\ell\sin{\phi},&&L_\text{b}=\dfrac{a}{L_\ell}\sec{\phi}-L_\ell\sin{\phi}.\label{eq:geom}
\end{align}
We nondimensionalise the single-cell energy~\eqref{eq:ecell} by scaling lengths with $\sqrt{a}$. The nondimensional energy is thus
\begin{widetext}
\vspace{-16pt}
\begin{align}
\hat{e}=\dfrac{e}{\sqrt{2a\Gamma\Gamma_\ell}}&=\dfrac{\sec{\phi}}{\Lambda}+\Lambda+\hat{B}\left[\biggl(\Lambda\sin{\phi}+\dfrac{\sec{\phi}}{\Lambda\ell_0^2}\biggr)^{-2}+\biggl(\Lambda\sin{\phi}-\dfrac{\sec{\phi}}{\Lambda\ell_0^2}\biggr)^{-2}+\dfrac{1}{\Lambda^2}\right],\label{eq:ecell2}
\end{align}
\vspace{-8pt}
\end{widetext}
where
\begin{align}
\ell_0^2&=\dfrac{2\Gamma}{\Gamma_\ell},&\Lambda&=\dfrac{L_\ell}{\sqrt{2a\Gamma/\Gamma_\ell}},&\hat{B}&=\dfrac{B}{\sqrt{8\Gamma^3a^3/\Gamma_\ell}}.\label{eq:params}
\end{align}
In what follows, we will drop the hats in Eqs.~\eqref{eq:ecell2} and~\eqref{eq:params} that signal our nondimensionalisation.

We note that $L_\ell$ and hence $\Lambda$ are the same for each cell in an epithelium in this minimal vertex model, so that the lateral sides of the isosceles cells match up.
\subsection{Single-cell shapes in a flat epithelium}
In a flat epithelium, cells are rectangular, $\phi=0$, with nondimensional height $L_\ell=\ell_0\Lambda$ and width $L_\text{a}=L_\text{b}=(\ell_0\Lambda)^{-1}$, from Eqs.~\eqref{eq:geom} and~\eqref{eq:params}. In what follows, we will take $\ell_0\gg1$ and $\Lambda=O(1)$. This is therefore the limit of columnar cells that also underpins previous studies of differential-tension models \cite{haas19,guharay25} and their biological motivation~\cite{hocevar12,storgel16}. The stretch~$\Lambda$ now minimises, from Eq.~\eqref{eq:ecell2},
\begin{align}
e_0=\dfrac{1}{\Lambda}+\Lambda+B\left(2\Lambda^2\ell_0^4+\dfrac{1}{\Lambda^2}\right).
\end{align}
For $B=0$, this is minimised at $\Lambda=1$, i.e., at $L_\ell=\ell_0$ and ${L_\text{a}=L_\text{b}=\smash{\ell_0^{-1}}}$. With cytoplasmic effects, $B>0$, the minimum is at $\Lambda=\lambda$, determined as a solution of $\partial e_0/\partial\Lambda=0$, i.e.,\looseness=-1
\begin{align}
4B\ell_0^4\lambda^4+\lambda^3-\lambda-2B=0.\label{eq:lambda}
\end{align}
In a flat epithelium of $N$ cells, each cell therefore has width $(\ell_0\lambda)^{-1}$, and so the epithelium has length ${L_0=N(\ell_0\lambda)^{-1}}$.
\subsection{Buckling problem}
As announced above, we study the buckling of an epithelial monolayer of $N$ cells under compression. The buckled configuration of the epithelium minimises the sum of the single-cell energies subject to the imposed compression, by a factor $D$ \figref{fig1}{c}. We describe the numerical implementation of this minimisation, based on Ref.~\cite{guharay25}, in detail in \citeappendix{appA}.

\begin{figure*}[t]
    \centering
    \includegraphics[width=\linewidth]{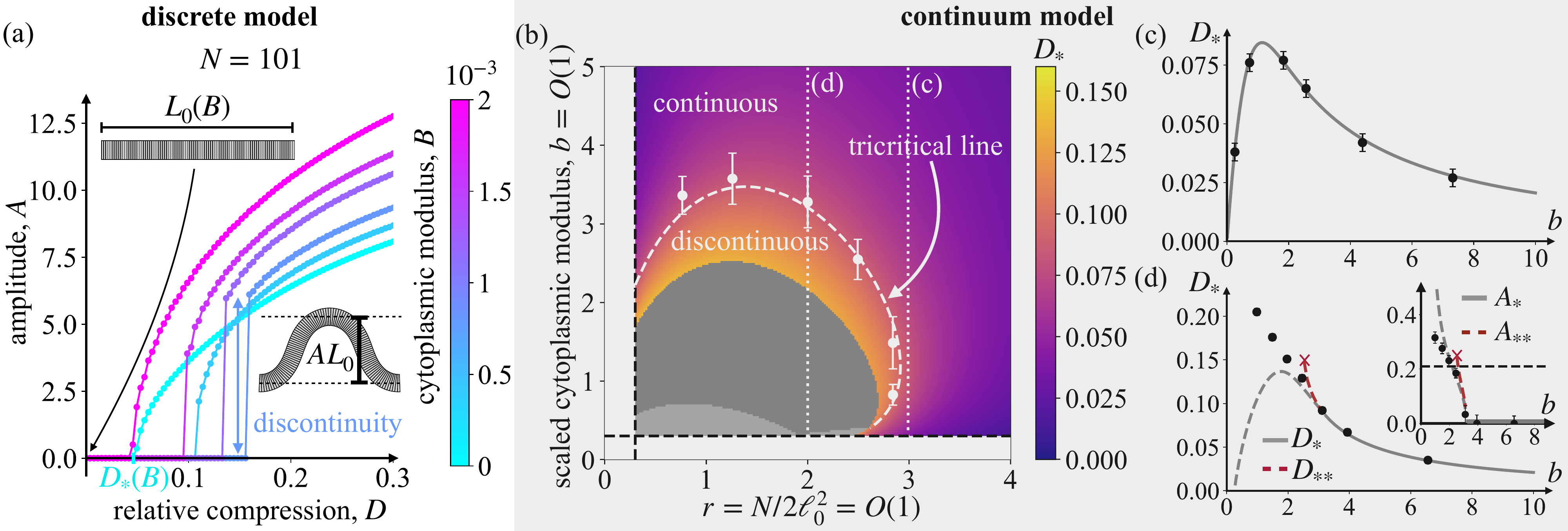}
    \caption{Continuous and discontinuous buckling of an epithelial monolayer under compression. (a) Numerical results for the discrete model. Plots of the buckling amplitude $A$ (inset) of a monolayer of $N$ cells against relative compression $D$, for different values of the cytoplasmic modulus $B$. The uncompressed length of the monolayer is $L_0(B)$ (inset). The buckling transition at $D=D_\ast(B)$ is continuous for small and large values of $B$ but discontinuous (arrow) in an intermediate range. Insets: flat and buckled shapes of the monolayer. Parameter values: $N = 101$, $\ell_0  = 4.5$. (b) Phase diagram of buckling in the continuum model. Heatmap of the buckling threshold $D_\ast$ in $(r,b)$ space, where $r=N/2\ell_0^2$ and $b = BN^2/4$. The asymptotic Landau theory determines a tricritical line that separates continuous from discontinuous buckling, in agreement with estimates from the discrete model (markers and error bars). In the grey regions, the asymptotic Landau analysis is not valid. See text for further explanation. (c) Plot of $D_\ast$ against $b$ at constant ``large'' $r$ [dotted line in panel (b)], showing non-monotonic behaviour. (d)~Similar plot at constant, smaller $r$ [dotted line in panel (b)] of $D_\ast$ and the buckling amplitude $D_{\ast\ast}$ for discontinuous buckling without noise (see text for further explanation). Inset: Corresponding plot of the buckling amplitudes $A_\ast,A_{\ast\ast}$ at $D = D_\ast,D_{\ast\ast}$, highlighting the transition between continuous buckling (solid line) and discontinuous buckling (dashed lines). The asymptotic theory has no solution for $D_{\ast\ast},A_{\ast\ast}$ beyond the point marked $\times$. The asymptotic results for $D_\ast,A_\ast$ agree with the discrete simulations (markers) while the asymptotic analysis is valid (``small'' $A_\ast$, horizontal line in the inset). Error bars smaller than marker size are not shown. Parameter values: (c) $r=2.99$, (d)~$r = 2$.}
    \label{fig2}
\end{figure*}

\section{\uppercase{Numerical results}}
We plot the amplitude $A$ of the configuration minimising the energy of the epithelium against its compression $D$ in \textfigref{fig2}{a}. At small values of $D$, the state of minimum energy is the flat configuration. At a critical compression $D_\ast$, $A$ increases and the epithelium buckles~\figref{fig2}{a}. For small values of the cytoplasmic modulus $B$, this increase of $A$ is continuous at ${D=D_\ast}$, as it is at $B=0$, in the model without cytoplasmic effects~\cite{haas19,guharay25}. Strikingly, it becomes discontinuous in an intermediate range of $B$. This behaviour is reentrant: at larger~$B$, buckling is, once again, continuous. We also observe that the buckling threshold $D_\ast$ first increases, then decreases with increasing $B$~\figref{fig2}{a}.

This shows that these cell-scale cytoplasmic mechanics can affect the buckling transition and hence tissue-scale mechanics even qualitatively. This is perhaps all the more surprising as instances of discontinuous buckling are few and far between: examples include the buckling of a rod within an elastic medium~\cite{savelev04} and the buckling of very thick beams~\cite{coulais15,chen20}.

To understand the mechanical basis for the transition between continuous and discontinuous buckling, we will develop an asymptotic Landau theory of buckling within the continuum limit of this discrete model~[Figs.~\figrefp{fig2}{b}--\figrefp{fig2}{d}]. We will first discuss more generally how to couch buckling within the framework of a Landau theory~\cite{landaulifshitz}.

\section{\uppercase{Landau theory of buckling}}\label{sec:landau}
In the language of statistical physics, continuous and discontinuous buckling are second-order and first-order phase transitions, respectively, in which the order parameter is the buckling amplitude $A$ \figrefi{fig2}{a}, and the relative compression $D$ is (an effective) temperature. This motivates describing buckling in terms of a Landau theory.

This analogy between buckling bifurcations and Landau theories of phase transitions, while natural, appears to have been made but rarely~\cite{savelev04,turzi20}, so we will review the Landau theories of second-order and first-order phase transitions below. We will focus on the cases in which the system is symmetric under a sign change of the order parameter, $A\leftrightarrow-A$, and hence the free energy has an expansion in even powers of $A$.

\subsection{Second-order transitions: fourth-order Landau theory}
A symmetric second-order phase transition is described by a fourth-order Landau theory, with free energy~\cite{landaulifshitz}

\begin{figure*}[t]
    \centering
    \includegraphics[width=1\linewidth]{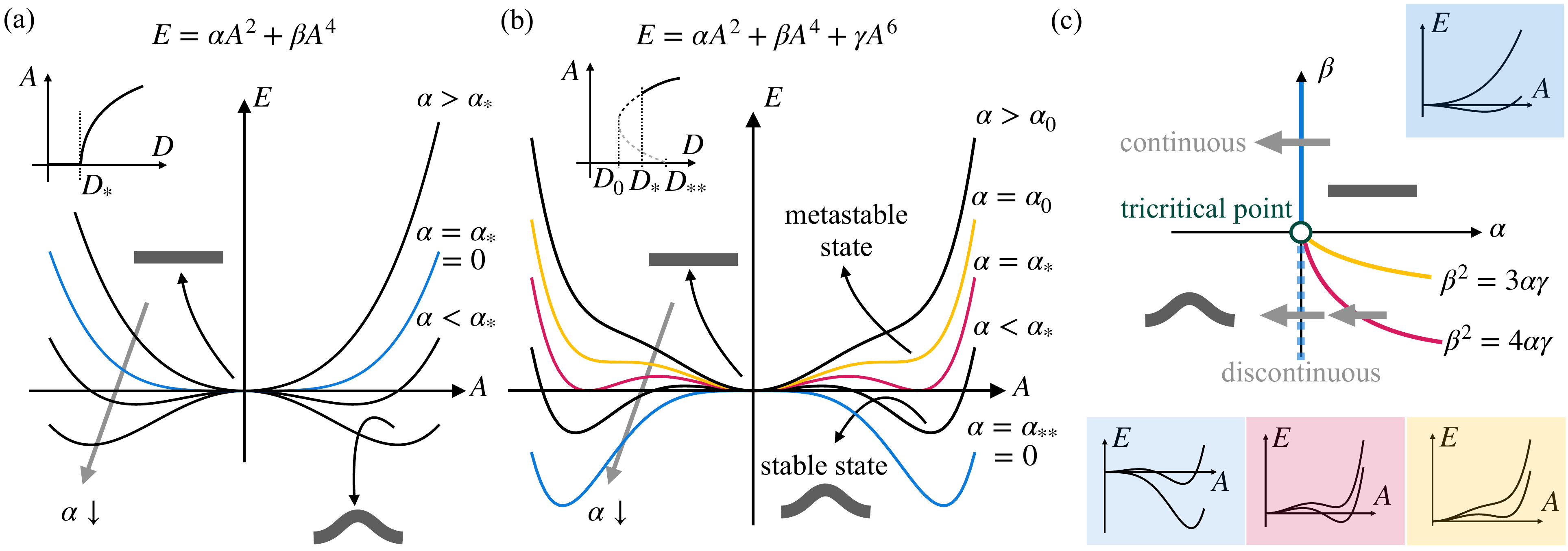}
    \caption{Review of Landau theories of second- and first-order phase transitions. (a) Fourth-order Landau theory of a second-order phase transition. Plot of the free energy $E = \alpha A^2 + \beta A^4$ against the order parameter $A$, for $\beta>0$.  At $\alpha = \alpha_\ast = 0$, the single minimum at $A=0$ for $\alpha>\alpha_\ast$ splits continuously into two symmetric minima at $A\neq 0$ for $\alpha<\alpha_\ast$. This results in a second-order phase transition (inset) as the control parameter $D$ and hence $\alpha$ vary. The bifurcation at $\alpha = \alpha_\ast$ corresponds to $D = D_\ast$. Shape insets: flat ($A=0$) and buckled ($A>0$) shapes. (b) Sixth-order Landau theory of a first-order phase transition. Plot of the free energy $E = \alpha A^2 + \beta A^4 + \gamma A^6$ against the order parameter $A$, for $\beta<0$, $\gamma>0$. For $\alpha>\alpha_0 = \beta^2/3\gamma$, there is a single minimum at $A=0$. At $\alpha = \alpha_0$, two metastable states appears as local minima with $A\neq 0$ and $E(A)>E(0)$. At $\alpha = \alpha_\ast = \beta^2 /4 \gamma$, these states become global minima. Finally, the minimum at $A=0$ disappears at $\alpha=\alpha_{\ast\ast}=0$. This discontinuous jump from $A=0$ to the global minima defines a first-order phase transition (inset). The critical points at $\alpha_0>\alpha_\ast>\alpha_{\ast\ast}$ correspond to $D_0<D_\ast<D_{\ast\ast}$, respectively. (c) Phase diagram of continuous and discontinuous buckling in $(\alpha,\beta)$ space, for $\gamma>0$. The continuous bifurcation happens at $\alpha=0$ for $\beta>0$. For $\beta<0$, the discontinuous bifurcation occurs at $\alpha=0$ in the absence of noise and at $\beta^2=4\alpha\gamma$ otherwise. Metastable solutions arise at $\beta^2=3\alpha\gamma$ and $\beta<0$. The bifurcation boundaries meet at the tricritical point $\alpha=\beta=0$. Insets illustrate the change in the $(A,E)$ energy landscape at these bifurcations. See text for further discussion.}
    \label{fig3}
\end{figure*}

\phantom{.}\vspace{-16pt}
\begin{align}
E=\alpha A^2+\beta A^4, \label{eq:L4}
\end{align}
with $\beta>0$. For $\alpha>\alpha_\ast=0$, $E$ has a single, global minimum at $A=0$ \figref{fig3}{a}. At $\alpha=\alpha_\ast$, this minimum splits into a maximum at $A=0$ and two symmetric minima at $A^2=|\alpha|/2\beta$ for $\alpha<\alpha_\ast$. In the buckling analogy, the latter correspond to the buckled states, while $A=0$ is the flat, compressed state. As $\alpha$ decreases across $\alpha_\ast=0$, $|A|$ increases continuously from $A=0$, defining a second-order transition and continuous buckling. The coefficients $\alpha,\beta$ are functions of the control parameter $D$, and this bifurcation corresponds to a critical value $D=D_\ast$ \figrefi{fig3}{a}. 

\subsection{First-order transitions: sixth-order Landau theory}
A symmetric first-order phase transition is described by a sixth-order Landau theory. The free energy is~\cite{landaulifshitz}
\begin{align}
E=\alpha A^2+\beta A^4+\gamma A^6, \label{eq:L6}
\end{align}
in which $\gamma>0$. If $\beta>0$, then Eq.~\eqref{eq:L6} recovers the fourth-order Landau theory \eqref{eq:L4} in the limit $A\ll 1$. In particular, there is still a second-order transition at $\alpha=\alpha_\ast=0$, so we turn to $\beta<0$. There is now a single minimum at $A=0$ for $\alpha>\alpha_0=\beta^2/3\gamma$ \figref{fig3}{b}. At $\alpha=\alpha_0$, two local minima appear at $|A|= (|\beta|/3\gamma)^{1/2}\neq 0 $. At this point, $E(A)>E(0)$, so these are metastable states~\figref{fig3}{b}. At $\alpha = \alpha_\ast = \beta^2/4\gamma$, these minima are at $|A| = (|\beta|/2\gamma)^{1/2}$ and $E(A) = E(0)$, so they become stable global minima for $\alpha<\alpha_\ast$. The minimum at $A=0$ is still a local minimum and remains separated from the global minima by a finite energy barrier. Finally, this energy barrier disappears at $\alpha = \alpha_{\ast \ast} = 0$, when $A = 0$ becomes a local maximum of $E$ \figref{fig3}{b}. At $\alpha= \alpha_{\ast \ast} $, the global minima are at $|A| = (2|\beta|/3\gamma)^{1/2}$. 

Thus, in the absence of fluctuations that can overcome the energy barrier between the local minimum at $A=0$ and the global minima, $A$ increases discontinuously from $A=0$ at $\alpha = \alpha_{\ast\ast}$. In the presence of fluctuations, this discontinuous increase of $A$ occurs earlier, when $\alpha = \alpha_\ast$, which defines the discontinuous phase transition of the Landau theory.

Again, the coefficients $\alpha, \beta,\gamma$ of the Landau theory are functions of $D$, so these bifurcations correspond to critical values $D=D_0<D_\ast< D_{\ast\ast}$, respectively \figrefi{fig3}{b}.

\subsection{Tricritical point}
We can assemble, in $(\alpha,\beta)$  space, these bifurcations into the diagram in \textfigref{fig3}{c}. In this diagram, the fourth-order Landau theory \eqref{eq:L4} occupies $\beta>0$ while the sixth-order Landau theory~\eqref{eq:L6} is represented in $\beta<0$. In particular, the continuous bifurcation is defined by the line $\alpha=0$ at $\beta>0$ while the discontinuous bifurcation occurs across $\beta^2 = 4 \alpha \gamma$ for $\beta<0$.

The bifurcation boundaries meet at the tricritical point ${\alpha = \beta = 0}$ which therefore marks the transition between continuous and discontinuous buckling.

\begin{figure}[t]
    \centering
    \includegraphics[width=8.6cm]{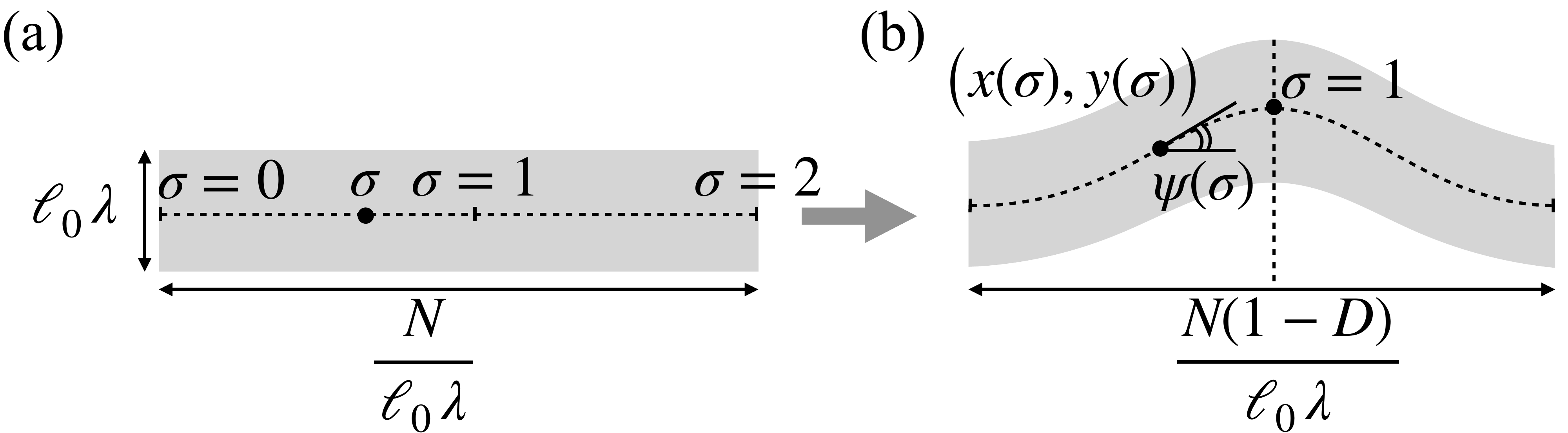}
    \caption{Continuum model. (a)~Uncompressed configuration: the monolayer has height $\ell_0\lambda$, length $N/(\ell_0\lambda)$, and is parameterised by relative arclength $\sigma\in[0,2]$. (b) Buckled configuration: under compression, the monolayer width becomes $N(1-D)/(\ell_0\lambda)$. In this configuration, the Cartesian coordinates of the point of the midline at relative arclength $\sigma$ in the undeformed configuration $\bigl(x(\sigma),y(\sigma)\bigr)$; the corresponding tangent angle is $\psi(\sigma)$.}
    \label{fig4}
\end{figure}

\section{\uppercase{Continuum theory}}
To explain the observed transition between continuous and discontinuous buckling \figref{fig2}{a}, we take the continuum limit of a monolayer of $N\gg 1$ cells. As we have already noted in our earlier discussion of single-cell shapes, the uncompressed monolayer has height $\ell_0\lambda$ and length $L_0=N/(\ell_0\lambda)$ \figref{fig4}{a}, where $\lambda$ solves Eq.~\eqref{eq:lambda}. Upon compression, this length becomes $L_0(1-D)=N(1-D)/(\ell_0\lambda)$ \figref{fig4}{b}. Let $\sigma\in[0,2]$ be a relative arclength along the uncompressed layer \figref{fig4}{a}. In the continuum limit, the single-cell energy of the discrete model, given by Eq.~\eqref{eq:ecell2}, becomes a function $e(\sigma)$ and the position of the midline of the monolayer is described by Cartesian coordinates $\bigl(x(\sigma),y(\sigma)\bigr)$, which define a tangent angle $\psi(\sigma)$ \figref{fig4}{b}. Clamping of the ends of the monolayer translates to $\psi(0)=\psi(2)=0$.

In line with previous work~\cite{haas19,guharay25}, we will focus on symmetric buckling \figref{fig4}{b}. This symmetry implies $\psi(\sigma)=-\psi(2-\sigma)$ for $\sigma\in[0,2]$; in particular $\psi(1)=0$. Hence the boundary conditions for the buckling of the half-layer $\sigma\in[0,1]$ are
\begin{align}
\psi(0)=\psi(1)=0.
\end{align}
The deformed monolayer is now described by $\psi(\sigma)$ and the lateral side length $\Lambda$ of the deformed cells, which we recall is constant along the monolayer, so that the lateral sides of the isosceles cells match up.

The shape of the deformed monolayer is determined by minimising the sum of the energies of the individual cells subject to the imposed relative compression, $D$. In the continuum limit, sums over cells are replaced by integrals over $\sigma$ according to
\begin{align}
\sum_{i=1}^{N}{}\to \dfrac{N}{2}\int_0^2{\mathrm{d}\sigma}=N\int_0^1{\mathrm{d}\sigma},
\end{align}
by symmetry. The nondimensional Lagrangian of the buckling problem is therefore
\begin{align}
\mathcal{L} &= N\int_0^1{\!\!e(\sigma)\,\mathrm{d}\sigma} + 2\mu\ell_0\left[x(1)\!-\!x(0)\!-\!\dfrac{N(1-D)}{2\ell_0\lambda}\right]\nonumber\\
&=N\int_0^1{\left\{e(\sigma)+\dfrac{2\mu\ell_0}{N}\left[x'(\sigma)-\dfrac{N(1-D)}{2\ell_0\lambda}\right]\right\}\,\mathrm{d}\sigma},\label{eq:Lagrangian}
\end{align}
where dashes denote differentiation with respect to $\sigma$ and in which the Lagrange multiplier $\mu$ is proportional to the compressive force on the monolayer (and where we have included a factor $2\ell_0$ for convenience). We can express the integrand in terms of $\psi(\sigma)$ using Eq.~\eqref{eq:ecell2} and the geometric relations (\citeappendix{appB})
\begin{subequations}\label{eq:geomrel}
\begin{align}
\phi(\sigma)&=\dfrac{\psi'(\sigma)}{N}+O\bigl(N^{-3}\bigr),\label{eq:geomrela}\\
x'(\sigma)&=\dfrac{N}{2\ell_0\Lambda}\left\{\left[1+\dfrac{\psi'(\sigma)^2}{6N^2}\right]\cos{\psi(\sigma)}-\dfrac{\psi''(\sigma)}{3N^2}\sin{\psi(\sigma)}\right\}\nonumber \\
&\qquad+O\bigl(N^{-3}\bigr).\label{eq:geomrelb}
\end{align}
\end{subequations}
We substitute Eqs.~\eqref{eq:geomrel} into Eq.~\eqref{eq:Lagrangian} to get an expansion of $\mathcal{L}$ in inverse powers of $N\gg 1$. However, we still need to choose the asymptotic scaling of the nondimensional cytoplasmic modulus $B$. We write $B=4b/N^2$, with a scaled cytoplasmic modulus $b=O(1)$. With this scaling,
\begin{align}
\dfrac{\mathcal{L}}{N}&=\int_0^1{\left\{\Lambda + \dfrac{1}{\Lambda} + \dfrac{8 b \Lambda^2\left[4r^2 +\Lambda^4\psi'(\sigma)^2\right]}{\left[4r^2-\Lambda^4\psi'(\sigma)^2\right]^2}\right.}\nonumber\\
&\hspace{34pt}+\mu\left[\dfrac{\cos{\psi(\sigma})}{\Lambda}-\dfrac{1-D}{\lambda}\right]\Biggr\}\,\mathrm{d}\sigma+O\bigl(N^{-2}\bigr),\label{eq:Lagrangian2}
\end{align}
in which, as in Ref.~\cite{guharay25}, we assume that $\smash{r=N/2\ell_0^2}$ is of order $O(1)$. From \textfigref{fig4}{a}, the aspect ratio of the undeformed monolayer is $\smash{N/\ell_0^2\lambda^2}$. Since $\Lambda,\lambda=O(1)$ in the limit of columnar cells that we are considering, this assumption therefore fixes the asymptotic scaling of this aspect ratio. 

Had we chosen a different asymptotic scaling for $B$, the first contribution from cytoplasmic mechanics [i.e., the third term in the integrand of Eq.~\eqref{eq:Lagrangian2}] would have appeared at a different order in the expansion. Our choice thus ensures that this contribution appears at the same order as the first contribution from cell surface mechanics, $\Lambda+1/\Lambda$, representing an effective stretching energy. This choice is justified \emph{a posteriori} by the fact that it explains the observed discontinuous buckling.

\subsection{Euler--Lagrange equations}
Varying Eq.~\eqref{eq:Lagrangian2} with respect to $\psi(\sigma),\Lambda,\mu$, we arrive at the Euler--Lagrange equation
\begin{align}
\frac{\mathrm{d}}{\mathrm{d}\sigma}\left(
\frac{12 r^2\psi' + \Lambda^4 \psi'^3}{\left(4 r^2 - \Lambda^4 \psi'^2\right)^3}
\right)+\frac{\mu \sin \psi}{16b\Lambda^7}=0,\label{eq:EL}
\end{align}
to be solved subject to the boundary conditions
\begin{subequations}\label{eq:ELbc}
\begin{align}
\psi(0)=\psi(1)=0\label{eq:ELbc1}
\end{align}
and the integral conditions
\begin{align}
&\int_0^1{\cos{\psi(\sigma)}\,\mathrm{d}\sigma} = \dfrac{\Lambda}{\lambda}(1-D),\label{eq:ELbc2}\\
&\int_0^1{\dfrac{16r^4 + 24 r^2 \Lambda^4 \psi'(\sigma)^2+\Lambda^8 \psi'(\sigma)^4}{\left[4r^2 - \Lambda^4 \psi'(\sigma)^2\right]^3}\,\mathrm{d}\sigma}\nonumber\\
&\quad=-\dfrac{1}{16b\Lambda}\left[1-\dfrac{1}{\Lambda^2}-\dfrac{\mu(1-D)}{\Lambda\lambda}\right],\label{eq:ELbc3}
\end{align}
\end{subequations}
in which we have used condition~\eqref{eq:ELbc2} to express the second condition in the form in Eq.~\eqref{eq:ELbc3}. In these conditions,~$\lambda$ is determined by Eq.~\eqref{eq:lambda}, which becomes
\begin{align}
\dfrac{4b}{r^2}\lambda^3+\lambda^2-1=0,\label{eq:lambda2}
\end{align}
up to corrections of order $O\bigl(N^{-2}\bigr)$, which are of the same order as those to the Lagrangian~\eqref{eq:Lagrangian2} neglected in the derivation of the Euler--Lagrange equations.
\subsection{Asymptotic Landau theory}
We derive our Landau theory by solving Eq.~\eqref{eq:EL} subject to Eqs.~\eqref{eq:ELbc} for a small amplitude $\epsilon\ll 1$ that we will relate to the physical amplitude $A$ below. We posit
\begin{subequations}
\begin{align}
\psi(\sigma) &= \psi_1(\sigma) \epsilon + \psi_3(\sigma) \epsilon^3 +\psi_5(\sigma)\epsilon^5+ O\bigl(\epsilon^7\bigr),\\
\Lambda &= \Lambda_0 + \Lambda_2 \epsilon^2 + \Lambda_4 \epsilon^4 + \Lambda_6\epsilon^6+O\bigl(\epsilon^8\bigr),\\
\mu &= \mu_0 + \mu_2 \epsilon^2 + \mu_4 \epsilon^4 + O\bigl(\epsilon^6\bigr).
\end{align}
\end{subequations}
The problem is symmetric under reflection in the undeformed midline. Under this reflection, ${\psi(\sigma)\leftrightarrow-\psi(\sigma)}$, but $\Lambda\leftrightarrow\Lambda$, $\mu\leftrightarrow\mu$. This is why $\psi(\sigma)$ has an expansion in odd powers of~$\epsilon$, while $\Lambda,\mu$ have expansions in even powers of $\epsilon$.
\subsubsection{Leading-order solution}
We start by considering the first integral condition, Eq.~\eqref{eq:ELbc2}, at order $O\bigl(\epsilon^0\bigr)$, which gives
\begin{align}
\Lambda_0=\dfrac{\lambda}{1-D}.
\end{align}
With this, Eqs.~\eqref{eq:EL} and~\eqref{eq:ELbc1} reduce at order $O\bigl(\epsilon^1\bigr)$ to
\begin{subequations}
\begin{align}
&\psi_1''(\sigma)+\dfrac{\mu_0 r^4(1-D)^7}{3b\lambda^7}\psi_1(\sigma)=0,&&\psi_1(0)=\psi_1(1)=0.
\end{align}
Requiring this problem to have a non-trivial solution, we conclude that the lowest buckling mode is
\begin{align}
&\psi_1(\sigma)=\sin{\pi\sigma},
\end{align}
for which
\begin{align}
    \mu_0=\dfrac{3\pi^2b\lambda^6}{r^4(1-D)^7}.
\end{align}
\end{subequations}
[More generally, the lowest mode is $\psi_1(\sigma)=a_1\sin{\pi\sigma}$, but the undetermined constant of integration $a_1$ can be absorbed into $\epsilon$.] Finally, Eq.~\eqref{eq:ELbc3} becomes, to leading order,
\begin{align}
1 + \dfrac{4 b \lambda}{r(1-D)} -\dfrac{(1-D)^2}{\lambda^2}-\dfrac{3\pi^2b}{\lambda^6(1-D)^5r^3} = 0, \label{eq:dstar}
\end{align}
where we recall that $\lambda=\lambda(b,r)$ is determined by Eq.~\eqref{eq:lambda2}. A solution $D=D_\ast$ of this equation determines the buckling threshold. We will use the full expansion of Eq.~\eqref{eq:ELbc3} below to obtain our asymptotic Landau theory.
\vspace{6pt}
\subsubsection{Next-order solution}
We continue the expansion to next order by evaluating condition~\eqref{eq:ELbc2} at order $O\bigl(\epsilon^2\bigr)$ to obtain
\begin{align}
\Lambda_2 = -\dfrac{\lambda}{4(1-D)}.
\end{align}
We return to Eqs.~\eqref{eq:EL}, now at order $O\bigl(\epsilon^3\bigr)$, to find, using \textsc{Mathematica} (Wolfram, Inc.),
\begin{widetext}
\vspace{-12pt}
\begin{subequations}
\begin{align}
\psi_3''(\sigma)+\pi^2\psi_3(\sigma)= \left[\dfrac{5 \pi^4 \lambda^4}{8 (1-D)^4 r^2}
-\dfrac{(1-D)^7 r^4 \mu_2}{3 b \lambda^7}-\dfrac{13\pi^2}{8}\right]\sin{\pi \sigma}  + \dfrac{\pi^2}{24}\left[\dfrac{15 \pi^2 \lambda^4}{(1-D)^4 r^2}-1\right]\sin{3\pi\sigma},
\end{align}
The solution of this differential equation is
\begin{align}
\psi_3(\sigma)&=a_3\sin{\pi\sigma}+b_3\cos{\pi\sigma}+\left[\dfrac{1}{192}
-\dfrac{5\pi^2\lambda^4}{64(1-D)^4 r^2}\right]\sin{3\pi \sigma}+\left[\dfrac{13\pi}{16}-\dfrac{5\pi^2\lambda^4}{16r^2(1-D)^4}+\dfrac{\mu_2r^4(1-D)^7}{6\pi b \lambda^7}\right]\sigma\cos{\pi\sigma},
\end{align}
\end{subequations}
in which $a_3,b_3$ are constants of integration. The first term can be absorbed into $\psi_1(\sigma)$ up to rescaling $\epsilon$. We may therefore set $a_3=0$ without loss of generality. Boundary conditions~\eqref{eq:ELbc1} then lead to
\begin{align}
&b_3=0,&&\mu_2 = \dfrac{3 \pi^2 b\lambda^7\left[5 \pi^2 \lambda^4-13(1-D)^4 r^2\right]}{8(1-D)^{11} r^6}.
\end{align}
\subsubsection{Solution at orders $O\bigl(\epsilon^4\bigr)$ and $O\bigl(\epsilon^5\bigr)$}
Unfortunately, we must proceed to still higher order in our asymptotic expansion to derive our Landau theory. Condition~\eqref{eq:ELbc2} at order $O\bigl(\epsilon^4\bigr)$ now gives\vspace{-10pt}
\begin{align}
\Lambda_4=\dfrac{\lambda}{64(1-D)}.
\end{align}
From Eq.~\eqref{eq:EL} at order $O\bigl(\epsilon^5\bigr)$, we then obtain
\begin{subequations}
\begin{align}
\psi_5''(\sigma)+\pi^2\psi_5(\sigma)&=\left\{\dfrac{\pi^2}{512}\left[621-\dfrac{830\pi^2\lambda^4}{(1-D)^4r^2}+\dfrac{65\pi^4\lambda^8}{(1-D)^8r^4}\right]-\dfrac{\mu_4(1-D)^7r^4}{3b\lambda^7}\right\}\sin{\pi\sigma}\nonumber\\
&\qquad -\frac{\pi^2\left[(1-D)^8 r^4 + 310(1-D)^4\pi^2 r^2\lambda^4 + 215\pi^4\lambda^8\right]}{512(1-D)^8 r^4}\sin{3\pi\sigma}\nonumber\\
&\qquad+\dfrac{\pi^2}{2560}\left\{\dfrac{25\pi^2\lambda^4}{(1-D)^8r^4}\left[6(1-D)^4r^2-61\pi^2\lambda^4\right]-3\right\}\sin{5\pi\sigma},
\end{align}
with solution
\begin{align}
\psi_5(\sigma)&=a_5\sin{\pi\sigma}+b_5\cos{\pi\sigma} 
- \left\{\dfrac{\pi \left[621 (1-D)^8 r^4 - 830 (1-D)^4 \pi^2 r^2 \lambda^4 + 65 \pi^4 \lambda^8\right]}
{1024 (1-D)^8 r^4}
-\frac{\mu_4(1-D)^7 r^4}
{6 b \pi \lambda^7}\right\}\sigma \cos{\pi\sigma} \nonumber\\
&\quad +
\dfrac{(1-D)^8 r^4 + 310 (1-D)^4 \pi^2 r^2 \lambda^4 + 215 \pi^4 \lambda^8}
{4096 (1-D)^8 r^4}\sin{3 \pi \sigma} + 
\dfrac{3 (1-D)^8 r^4 - 150 (1-D)^4 \pi^2 r^2 \lambda^4 + 1525 \pi^4 \lambda^8}
{61440 (1-D)^8 r^4}
\sin{5\pi\sigma}
, 
\end{align}
\end{subequations}
in which $a_5,b_5$ are more constants of integration. Again, absorbing the first term into $\psi_1(\sigma)$, we can set $a_5=0$, and then use boundary conditions~\eqref{eq:ELbc1} to obtain\newpage

\phantom{.}\vspace{-18pt}
\begin{align}
&b_5=0,&&\mu_4=
\dfrac{3 b \pi^2 \lambda^7 \left[621 (1-D)^8 r^4 - 830 (1-D)^4 \pi^2 r^2 \lambda^4 + 65 \pi^4 \lambda^8\right]}
{512 (1-D)^{15} r^8}.
\end{align}
\subsubsection{Solution at order $O\bigl(\epsilon^6\bigr)$}
We need one more result before computing our asymptotic Landau theory: the final step of the asymptotic expansion is to evaluate condition~\eqref{eq:ELbc2} at order $O\bigl(\epsilon^6\bigr)$, which yields
\begin{align}
\Lambda_6=-\dfrac{\lambda\left[3(1-D)^4r^2-5\pi^2\lambda^4\right]^2}{16384(1-D)^9r^4}.
\end{align}
\subsubsection{Calculation of the buckling amplitude}
Finally, we need to relate the physical buckling amplitude $A$ of the midline \figrefi{fig2}{a} to the asymptotic parameter $\epsilon$. For this purpose, we use the geometric relation [\textfigref{fig4}{b} and \citeappendix{appB}]
\begin{align}
y'(\sigma)=\dfrac{N}{2\ell_0\Lambda}\left\{\left[1+\dfrac{\psi'(\sigma)^2}{6N^2}\right]\sin{\psi(\sigma)}+\dfrac{\psi''(\sigma)}{3N^2}\cos{\psi(\sigma)}\right\}+O\bigl(N^{-3}\bigr).\label{eq:ygeom}
\end{align}
This yields, using the expansions of $\mu$, $\Lambda$, and $\psi(\sigma)$ computed above,
\begin{subequations}
\begin{align}
A=\dfrac{y(1)-y(0)}{L_0}=\int_0^1{\dfrac{y'(\sigma)}{L_0}\,\mathrm{d}\sigma}&=\left[\dfrac{1-D}{\pi}\epsilon + \dfrac{27(1-D)^4 r^2 - 5 \pi^2 \lambda^4}{192(1-D)^3 \pi r^2} \epsilon^3 +\dfrac{75(1-D)^8 r^4 + 40(1-D)^4 \pi^2 r^2 \lambda^4 + 69\pi^4\lambda^8}{3072(1-D)^7\pi r^4}\epsilon^5\right.\nonumber\\
&\qquad+O\bigl(\epsilon^7\bigr)\biggr]+O\bigl(N^{-2}\bigr),\label{eq:midlineA}
\end{align}
where $L_0 = N/\ell_0\lambda$ is the length of the uncompressed monolayer \figrefi{fig2}{a}. The corrections at order $O\bigl(N^{-2}\bigr)$ are of the same order as the terms neglected in the expansion~\eqref{eq:Lagrangian2} of the Lagrangian, so we can ignore them in what follows. Inverting this expansion order-by-order gives
\begin{align}
\epsilon=  \frac{\pi}{1-D} A - \frac{\pi^3 \left[27(1-D)^4r^2-5\pi^2\lambda^4\right]}{192(1-D)^7r^2} A^3 + \frac{\pi^5
\left[429(1-D)^8r^4-430(1-D)^4\pi^2r^2\lambda^4-251\pi^4\lambda^8\right]}{12288(1-D)^{13}r^4} A^5+O\bigl(A^7\bigr).
\end{align}
\end{subequations}
\subsubsection{Derivation of the asymptotic Landau theory}
Finally, we return to the integral condition~\eqref{eq:ELbc3}, which becomes $\hat\alpha+\hat\beta A^2+\hat\gamma A^4=O\bigl(A^6\bigr)$, in which $\hat\alpha,\hat\beta,\hat\gamma$ are (complicated) functions of $b,r,\lambda,D$. Neglecting the term at order $O\bigl(A^6\bigr)$, this equation determines the stationary points of
\begin{align}
E=\alpha A^2+\beta A^4+\gamma A^6,\label{eq:asymptoticLandau}
\end{align}
with $\alpha=\hat\alpha/2$, $\beta=\hat\beta/4$, $\gamma=\hat\gamma/6$. Equation~\eqref{eq:asymptoticLandau} thus defines our asymptotic Landau theory. We emphasise that it results from a double asymptotic expansion of the Lagrangian of the buckling problem, for $A\ll 1$ and $N\gg 1$. Explicitly, the coefficients are, up to a positive overall factor that is of no consequence,
\begin{subequations}\label{eq:landaucoeff}
\begin{align}
\alpha & = 3 b \pi^2 \lambda^7 + (1-D)^5 r^4\left[(1-D)^2 - \lambda^2\right]-4b(1-D)^4r^2 \lambda^3 \\
\beta &= 
\dfrac{
8 b (1-D)^8 r^4 \lambda^3
- 69 b (1-D)^4 \pi^2 r^2 \lambda^7
+15 b \pi^4 \lambda^{11}
+4 (1-D)^{11} r^6}
{16(1-D)^{4} r^2},\\
\gamma&=\dfrac{80 (1-D)^{15} r^8 - 32 b (1-D)^{12} r^6 \lambda^3 + 4191 b (1-D)^8 \pi^2 r^4 \lambda^7 - 3450 b (1-D)^4 \pi^4 r^2 \lambda^{11} + 195 b \pi^6 \lambda^{15}}{1536(1-D)^8 r^4}.
\end{align}
\end{subequations}
\subsubsection{Computing the buckling transitions}
We compute the buckling transitions from Eqs.~\eqref{eq:landaucoeff} and the general results obtained in Sec.~\ref{sec:landau}. In particular, the conditions $\alpha=0$ for a second-order transition and $\beta^2=4\alpha\gamma$ for a first-order transition define polynomial equations for $D=D_\ast$. For these values of $D_\ast$, we check that the additional conditions $\beta>0$ for a second-order transition and $\beta<0,\gamma>0$ for a first-order transition hold. We report the transition associated with the smallest $D_\ast$ found in this way, and also compute the associated amplitude $A_\ast$ using the general results of Sec.~\ref{sec:landau}. Similarly, we compute $D_{\ast\ast},A_{\ast\ast}$.
In particular, tricritical points satisfy $\alpha=\beta=0$ with $\gamma>0$. This means that the polynomial numerators of $\alpha,\beta$ have a double root in $D$, so their resultant~\cite{gelfand} 
\begin{align}
\mathcal{R}&\propto 
12098945286144 b^{10} 
- 150000 \pi^{4} r^{16}- 27 b^{6} \bigl(
164935095243 \pi^{6} r^{2}
- 196849422304 \pi^{4} r^{4}
+ 28941571392 \pi^{2} r^{6}
\bigr)
\nonumber\\
&\qquad
+ 3786752 b^{8} \bigl(5415849 \pi^{2} r^{2} - 29584 r^{4}\bigr)
+ 1125 b^{2} \pi^{4} r^{4} \bigl(
1875 \pi^{8}
- 33750 \pi^{6} r^{2}
+ 151575 \pi^{4} r^{4}
- 123360 \pi^{2} r^{6}
+ 56768 r^{8}
\bigr)\nonumber\\
&\qquad- 72 b^{4} r^4 \bigl(
484128375 \pi^{8} 
- 569625750 \pi^{6} r^{2}
+ 71391740 \pi^{4} r^{4}
- 74270282 \pi^{2} r^{6}
\bigr)\label{eq:res}
\end{align}\newpage
\end{widetext}
vanishes. The proportionality sign in Eq.~\eqref{eq:res} hides a nonzero factor of no consequence. This determines a line of tricritical points in the $(r,b)$ plane, which we plot together with $D_\ast$ in \textfigref{fig2}{b} for $r,b=O(1)$. Moreover, we plot slices of this diagram at constant $r$ in Figs.~\figrefp{fig2}{c} and~\figrefp{fig2}{d}. This diagram excludes regions of parameter space in which (i) the asymptotic Landau theory breaks down [dark grey region in \textfigref{fig2}{b}] because $A_\ast\not\ll 1$ [as indicated by the horizontal line in the inset of \textfigref{fig2}{d}] or (ii) this Landau theory does not predict a buckling transition of either order [light grey region in \textfigref{fig2}{b}]. In the latter case, continuous or discontinuous buckling may be described by a higher-order Landau theory. Following up on this possibility is beyond the scope of our paper.
\subsection{Discussion}
The continuum theory reproduces the features of the discrete model observed in \textfigref{fig2}{a}: the tricritical line has a fold point, $\partial r/\partial b=0$ \figref{fig2}{b}, so, for ``small'' values of $r$, the buckling transition is discontinuous in an intermediate range of $b$ only, in agreement with the results for the discrete model in \textfigref{fig2}{a}. There is another fold, $\partial b/\partial r=0$, again in agreement with the simulations. Thus the discontinuous regime can be reentrant both as $r$ is varied and as $b$ is varied. For ``large'' values of $r$ or ``large'' values of $b$, the buckling transition is continuous for all $b$ or all $r$, respectively \figref{fig2}{b}; this is again consistent with the discrete model (not shown).

This comparison between the discrete model and the continuum theory can also be made more quantitative: numerical estimates of tricritical points for the discrete model fall onto the tricritical line computed from our asymptotic Landau theory \figref{fig2}{b}. Similarly, estimates of $D_\ast$ and $\mathcal{A}_\ast$ for the discrete model agree with the asymptotic results [Figs.~\figrefp{fig2}{c} and~\figrefp{fig2}{d}] where the asymptotic analysis is valid (i.e., where $A_\ast\ll 1$).

Our continuum theory thus shows that discontinuous buckling results from the competition between the stretching energy density of the cell cortices [corresponding to $\Lambda+1/\Lambda$ in Eq.~\eqref{eq:Lagrangian2}] and a contribution from cytoplasmic mechanics (i.e., the third term in this integrand). Expanding the latter for small amplitudes, $\psi'(\sigma)\ll 1$, gives
\begin{align}
\dfrac{8 b \Lambda^2\left[4r^2 +\Lambda^4\psi'(\sigma)^2\right]}{\left[4r^2-\Lambda^4\psi'(\sigma)^2\right]^2}\sim\dfrac{2b\Lambda^2}{r^2}+\dfrac{3b\Lambda^6}{2r^4}\psi'(\sigma)^2.
\end{align}
The first term modifies the stretching energy; the second is proportional to curvature squared, so represents an emergent bending energy. Thus the separation of stretching and bending energies that one expects to drive buckling persists. However, a different bending energy, $\psi'(\sigma)^2/2\Lambda N^2$, arises from cell surface mechanics at order $O\bigl(N^{-2}\bigr)$ in the expansion of Eq.~\eqref{eq:Lagrangian2}. It is this term that controls buckling at $B=0$ \cite{haas19,guharay25}.

\section{\uppercase{Conclusion}}
In this paper, we have studied the buckling under compression of an epithelial layer in a minimal vertex model that includes a mechanical contribution from bulk, cytoplasmic effects. We have discovered an unusual discontinuous buckling transition, complementing the few known instances of discontinuous buckling~\cite{savelev04,coulais15,chen20}, and we have explained it using coarse-grained continuum theory.

Our work thus shows more generally how adding additional mechanobiological effects at the cell scale to vertex models can change their emergent, tissue-scale behaviour even qualitatively. This contrasts to some extent with the universal behaviour observed in vertex models in other contexts~\cite{damavandi25}.

The picture of a sixth-order Landau theory and its tricritical points has provided us with a useful framework to understand especially the discontinuous buckling transition. We reiterate that we are not the first to make this connection between Landau theories and mechanical bifurcations~\cite{savelev04,turzi20}, but it remains a bit surprising to us that this analogy does not appear to have been drawn more often. This analogy does raise however a more abstract question: what mechanical systems realise more complex bifurcations that are described by Landau theories of higher order? Conversely, the Landau theory framework is, more generally, a tool to describe the effect of fluctuations near phase transitions~\cite{landaulifshitz}. Can this analogy be developed to understand the effect of fluctuations in mechanical problems?\looseness=-1

The assumption of isosceles cell shapes in our minimal vertex model has allowed us to understand the discontinuous buckling transition analytically. Nevertheless, the phenomenology of discontinuous buckling is not an artefact of this assumption: in \citeappendix{appC}, we show that a generalised model with general quadrilateral cell shapes also displays discontinuous buckling. Moreover and as already noted in the Introduction, we reported unbuckling in Ref.~\cite{guharay25}: in the absence of cytoplasmic mechanics (i.e., in our minimal vertex model with $B=0$), the buckling amplitude can decrease with increasing compression. In \citeappendix{appD}, we show that unbuckling persists for $B>0$. This stresses that these phenomena, of discontinuous buckling and unbuckling, are robust.

We have focused on a particular mechanobiological contribution of the cytoplasm, the cost of confining cytoplasmic polymers that has previously been included in the vertex model of Ref.~\cite{hannezo14}. In our model, this cost diverges with the length scale $L$ of the confinement like $L^{-2}$~\cite{degennes}. This divergence makes it a more extreme contribution than more generic bulk contributions that have been included in vertex models, such as volume elasticity~\cite{honda04,hannezo14,zhang22,villeneuve24,sarkar26}. By imposing incompressibility, we have taken the limit of an infinite modulus of volume elasticity. Still, there are other possible mechanobiological contributions: for example, the high stiffness of the nucleus~\cite{lammerding11,kalukula22} may start to dominate the mechanics beyond a critical compression, giving rise to effective strain-stiffening behaviour. This effect may be significant in pseudostratified epithelia~\cite{norden17,gomez21}: for example, zebrafish retinal morphogenesis involves a nuclear jamming transition~\cite{ferme25}. Its mechanical consequences for even the simple buckling instability that we have analysed here remain to be understood fully. 

Finally, buckling instabilities pervade the emergence of biological shape during development~\cite{nelson16}. They include both instabilities \emph{of} surfaces, driving, e.g., brain folding~\cite{tallinen14,budday15,budday15b,tallinen16,budday18}, but also instabilities \emph{within} surfaces~\cite{zhao25}, e.g., during \emph{Drosophila} gastrulation~\cite{smits23,alber25,serafini26}. The simple buckling problem that we have studied in this paper is therefore not merely an analytically solvable toy problem, but also a minimal model that determines the role of cytoplasmic effects in these biological instabilities.

Understanding the emergent contribution of these cytoplasmic effects to tissue mechanics thus remains, more broadly, a key step to be taken towards quantitative, rather than merely qualitative mechanical models of tissues. 

\begin{acknowledgments}
We thank an anonymous reviewer of Ref.~\cite{guharay25} for asking a question about the effect of cytosolic mechanics on the unbuckling bifurcation that prompted this work. We also thank Pawel Korzeb for discussions and comments on the manuscript. We are grateful for funding from the Max Planck Society.
\end{acknowledgments}

\section*{Data availability}
Code for the numerical simulations of the vertex model is openly available at Ref.~\footnote{Code is available at \href{https://doi.org/10.5281/zenodo.23102926}{\texttt{doi.org/10.5281/zenodo.23102926}}}.

\appendix
\titleformat{\section}{\normalfont\bfseries\centering\small}{APPENDIX \thesection:}{0.5em}{}
\section{\uppercase{Numerical implementation of the buckling problem}}\label{appA}
In this Appendix, based on Ref.~\cite{guharay25}, we describe the numerical implementation of the buckling of an epithelial monolayer of $N$ discrete cells with clamped ends due to an imposed relative compression $D$.

The symmetric buckling problem can be reduced to minimising the energy of (the left) half of the monolayer, but the details then depend on whether $N$ is even or odd~\cite{guharay25}. Here, for simplicity, we will focus on the odd case, $N=2n-1$. The total energy of the monolayer is then
\begin{align}
\mathcal{E}&=\sum_{i=2}^{n-1}{\left(\dfrac{\sec{\phi_i}}{\Lambda}+\Lambda\right)}+\dfrac{1}{2}\left(\dfrac{\sec{\phi_1}+\sec{\phi_n}}{\Lambda}+2\Lambda\right) + B\dfrac{n-1}{\Lambda^2}\nonumber\\
&\quad+\sum_{i=2}^{n-1} B\left[\left(\Lambda \sin{\phi_i} + \dfrac{\sec{\phi_i}}{\Lambda \ell_0^2}\right)^{-2}+\left(\Lambda \sin{\phi_i} - \dfrac{\sec{\phi_i}}{\Lambda \ell_0^2}\right)^{-2}\right] \nonumber\\
&\quad+\dfrac{B}{2}\left[\left(\Lambda \sin{\phi_1} + \dfrac{\sec{\phi_1}}{\Lambda \ell_0^2}\right)^{-2}+\left(\Lambda \sin{\phi_1} - \dfrac{\sec{\phi_1}}{\Lambda \ell_0^2}\right)^{-2}\right.\nonumber \\ 
&\quad\hspace{8mm}+\left.\left(\Lambda \sin{\phi_n} + \dfrac{\sec{\phi_n}}{\Lambda\ell_0^2}\right)^{-2}\!\!+\left(\Lambda \sin{\phi_n} - \dfrac{\sec{\phi_n}}{\Lambda \ell_0^2}\right)^{-2}\right],
\label{eq:halfe}
\end{align}
where $2\phi_i$ is now the angle that the lateral sides of cell $i$ make with each other, for $i=1,2,\dots,n$. Again, the stretch $\Lambda$ of the lateral sides is a constant (i.e., the same for all cells) because adjacent cells have to match up. In this expression, the leftmost cell $1$ and the centre cell $n$ only contribute half the cell energy, as in Ref.~\cite{guharay25}; for the leftmost cell, this imposes the clamping of the monolayer.

Let $\psi_i$ now denote the angle that cell $i$ makes with the horizontal, for $i=1,2,\dots,n$. The clamped boundary conditions and the symmetry imply $\psi_1=\psi_n = 0$. We impose this condition using the recursive relation~\cite{haas19}
\begin{align}
\psi_i = \psi_{i-1} + \phi_i + \phi_{i-1},\quad\text{for }=1,2,\dots,n.
\end{align}
The horizontal position $x_i$ of the centre of cell $i$ satisfies the recursion~\cite{haas19}
\begin{align}
x_i = x_{i-1} + \frac{\sec{\phi_i}\cos{\psi_i} + \sec{\phi_{i-1}} \cos{\psi_{i-1}}}{2\ell_0\Lambda},
\end{align}
for $i=1,2,\dots,n$, with $x_1=0$. The imposed compression of the monolayer is expressed by the constraint~\cite{guharay25}
\begin{align}
x_n=\dfrac{n-1}{\ell_0}(1-D).
\end{align}
We minimise Eq.~\eqref{eq:halfe} subject to these constraints over the angles $\{\phi_1,\phi_2,\dots,\phi_n\}$ and determine the buckled shape of the monolayer using the \textsc{python3} optimisation routine \texttt{scipy.optimize}~\cite{scipy}.

\begin{figure}[b]
    \centering
    \includegraphics[width=8.6cm]{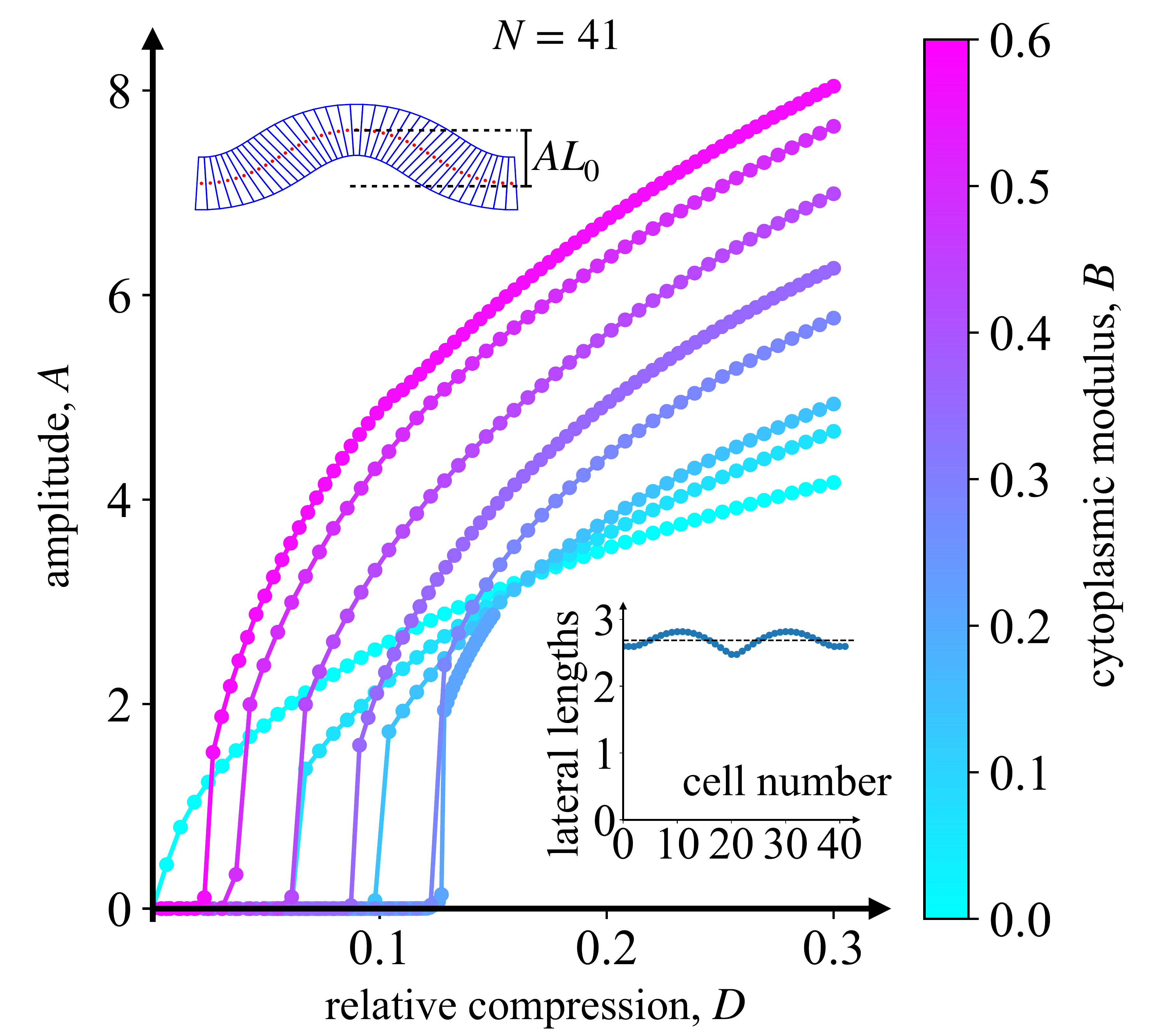}
    \caption{Discontinuous buckling of a compressed monolayer of $N$ cells in a generalised vertex model. Plot of the amplitude $A$ (defined in the shape inset) against relative compression $D$, for different values of the cytoplasmic modulus $B$. Inset: variation of the lateral cell sides along the monolayer for the buckled shape shown in the shape inset. Parameter values: $N=41$, $\ell_0=3$.}
    \label{fig5}
\end{figure}

\section{\uppercase{Derivation of Eqs.}~\eqref{eq:geomrel} AND~\eqref{eq:ygeom}}\label{appB}
Reference~\cite{haas19} has expressed the angle $\phi$ of the lateral cell sides and the positions $(x,y)$ of the midline in terms of the tangent angle $\psi$ using the discrete relations
\begin{subequations}
\begin{align}
\psi_{i+1}-\psi_i&=\phi_{i+1}+\phi_i,\label{eq:psidisc}\\
x_{i+1}-x_i&=\dfrac{1}{2\ell_0\Lambda}\left[\dfrac{\cos{\psi_{i+1}}}{\cos{\phi_{i+1}}}+\dfrac{\cos{\psi_i}}{\cos{\phi_i}}\right],\label{eq:xdisc}\\
y_{i+1}-y_i&=\dfrac{1}{2\ell_0\Lambda}\left[\dfrac{\sin{\psi_{i+1}}}{\cos{\phi_{i+1}}}+\dfrac{\sin{\psi_i}}{\cos{\phi_i}}\right],\label{eq:ydisc}
\end{align}
\end{subequations}
in which the index $i$ runs over the cells. In the continuum limit and in our notation, if index $i$ corresponds to relative arclength~$\sigma$, then index $i+1$ corresponds to $\sigma+1/N$. Taylor expanding Eq.~\eqref{eq:psidisc} and inverting order-by-order gives
\begin{align}
\phi(\sigma)=\dfrac{\psi'(\sigma)}{N}-\dfrac{\psi'''(\sigma)}{3N^3}+O\bigl(N^{-5}\bigr),
\end{align}
similarly to Ref.~\cite{haas19}. This implies Eq.~\eqref{eq:geomrela}. With this result, as in Ref.~\cite{haas19}, Taylor expanding Eqs.~\eqref{eq:xdisc} and~\eqref{eq:ydisc} leads to Eqs.~\eqref{eq:geomrelb} and~\eqref{eq:ygeom}.

\section{\uppercase{Discontinuous buckling in a generalised vertex model}}\label{appC}
In order to show that the discontinuous buckling transition is not an artefact of the assumption of isosceles cell shapes in our minimal vertex model, we perform simulations of a generalised vertex model in which the cell shapes are generic quadrilaterals. The description and numerical implementation of a generalised discrete model of this kind has been described in Sec.~II of the Supplemental Material of \citeref{guharay25}. Here, we add a cytoplasmic contribution to this model, so that the energy of the single cell $i$ [Eq.~(S11) of this Supplemental Material] is replaced with
\begin{align}
e_i = \dfrac{\ell_0^2}{2}\bigl(\ell_i^+ + \ell_i^-\bigr) + \ell_i + B\left[\dfrac{1}{(\ell_i^+)^2} + \dfrac{1}{(\ell_i^-)^2} + \dfrac{1}{\ell_i^2} \right].
\end{align}
Figure~\ref{fig5} illustrates discontinuous buckling in this generalised vertex model. The cell shapes of the buckled monolayer do not remain isosceles (\textwholefigref{fig5}, inset).

\section{\uppercase{Unbuckling bifurcation in the minimal vertex model}}\label{appD}
\begin{figure}[t]
    \centering
    \includegraphics[width=8.6cm]{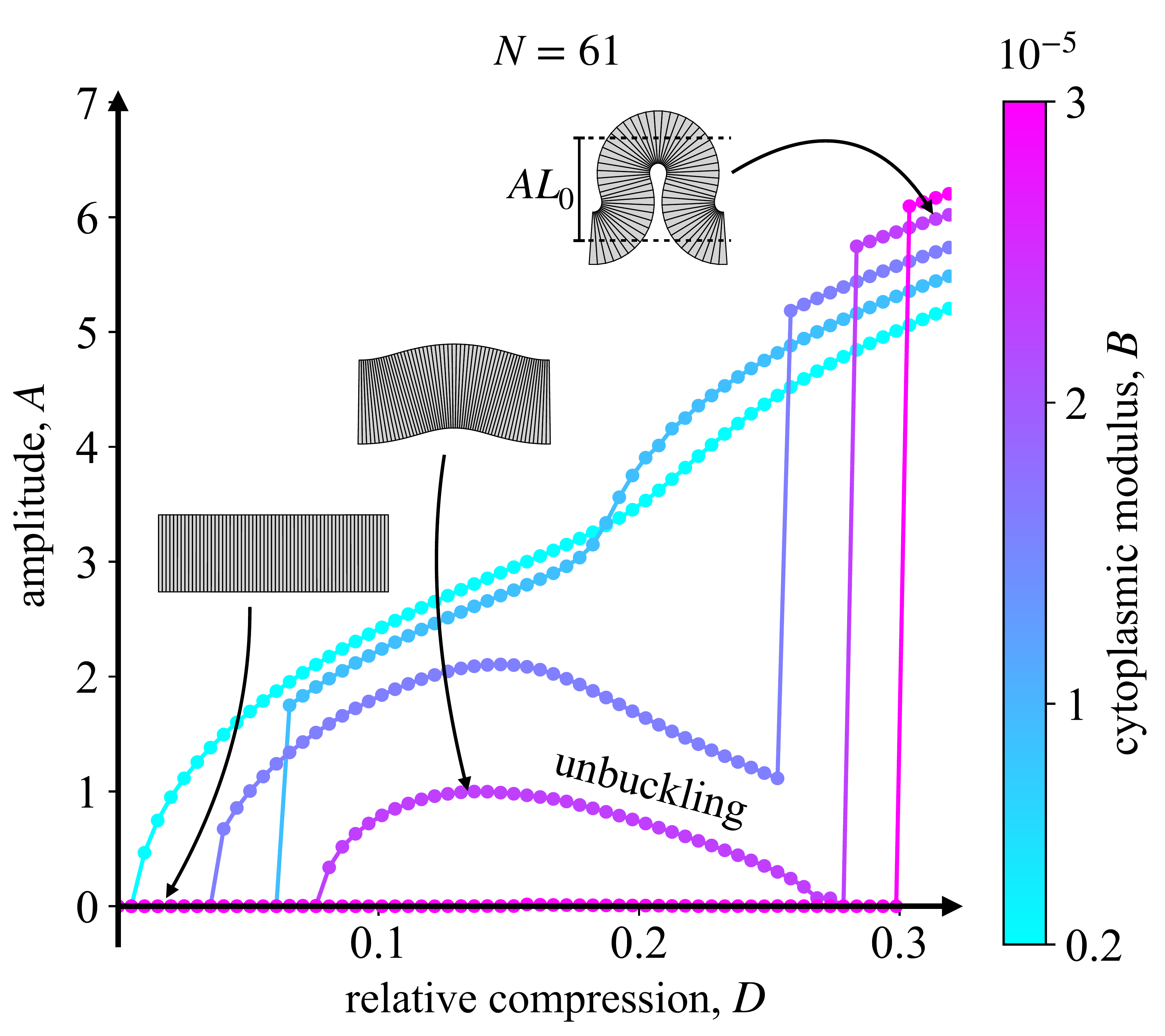}
    \caption{Unbuckling of a compressed monolayer of $N$ cells for $B>0$ in our minimal vertex model. Plot of the amplitude $A$ (defined in the shape inset) against relative compression $D$, for different values of the cytoplasmic modulus $B$. Parameter values: $N=61$, $\ell_0=4.5$.}
    \label{fig6}
\end{figure}

In order to demonstrate that unbuckling, as observed in \citeref{guharay25}, is a robust phenomenon, we show that it persists in our minimal model with cytoplasmic effects at $B>0$. Figure~\ref{fig6} shows unbuckling with varying cytoplasmic modulus~$B$. The buckling discontinuity at compressions $D$ below or above the unbuckling regime~\wholefigref{fig6}. Finally, we note that, as in Ref.~\cite{guharay25}, unbuckling arises for shorter, squat monolayers: \textwholefigref{fig6} shows results for $N=61$ cells, compared to the value $N=101$ used in \textfigref{fig2}{a}.
\newpage
\phantom{.}\vspace{-32pt}

\bibliography{main}
\end{document}